\documentclass[usenatbib]{mn2e}
\usepackage{setspace}
\input{psfig}
\newcommand{\dd}{{\rm d}}

\newcommand{\macc}{M_{\rm acc}}
\newcommand{\mh}{M_{\rm h}}

\newcommand{\etal}{{et al.}}

\newcommand{\beq}{\begin{equation}}
\newcommand{\eeq}{\end{equation}}

\newcommand{\mpch}{\>h^{-1}{\rm {Mpc}}}

\newcommand{\Om}{\Omega_{\rm M}}
\newcommand{\Odm}{\Omega_{\rm dm}}
\newcommand{\Obaryon}{{\Omega_{\rm b}}}

\newcommand{\seight}{\sigma_8}

\newcommand{\msunh}{\>h^{-1}\rm M_\odot}

\newcommand{\apj}{ApJ}
\newcommand{\apjs}{ApJS}

\newcommand{\mnras}{MNRAS}

\newdimen\hssize
\newdimen\hdsize 
\def\ltsima{$\; \buildrel < \over \sim \;$}
\def\lsim{\lower.5ex\hbox{\ltsima}}
\def\gtsima{$\; \buildrel > \over \sim \;$}
\def\gsim{\lower.5ex\hbox{\gtsima}}

\begin{document}


\title[Mass distribution and accretion of sub-halos]
      {Mass Distribution and Accretion of Sub-halos}
\author[Li, Mo]
       {Yun~Li$^1$\thanks{E-mail: liyun@astro.umass.edu}, H.J.~Mo$^1$\\
        $^1$Department of Astronomy, University of Massachusetts, 
        Amherst, MA 01003, USA\\
	}


\date{}

\pagerange{\pageref{firstpage}--\pageref{lastpage}}
\pubyear{2000}

\maketitle

\label{firstpage}


\begin{abstract}
We use the ``Millennium Simulation'' to study the mass function of accreted 
sub-halos during merger events in the dark halo assembly history. 
Our study includes three kinds of sub-halo mergers:
(1) mergers that happen to the main progenitor of dark halos;
(2) mergers that happen on the entire merging history tree of dark halos; 
and (3) mergers that leave {\it identifiable} sub-halos in present-day 
dark halos. We estimate the {\it unevolved} 
sub-halo mass functions (USMFs), for which sub-halo masses are 
measured at the times of their accretion. For sub-halos that merge into 
the main branch of a present-day dark halo, their USMF can be well 
described by a universal functional form, in excellent agreement 
with previous results. The same conclusion can also be reached for the 
USMF of all progenitors that have merged to become sub-halos during the 
entire halo merging history. In both cases, the USMFs are also 
independent of the redshift of host halos. Due to tidal disruption,
only a small fraction of the accreted halos survive as sub-halos
identifiable in the present-day dark halos. In cluster-sized halos, 
about 30\% of the survived sub-halos are sub-subhalos, and this 
fraction decreases with decreasing halo mass. For given halo and 
sub-halo masses, the accretion time has very broad distribution, but 
the survived sub-halos are all accreted quite recently.    
\end{abstract}


\section{Introduction}
\label{sec:intro}

In the past two decades, the Cold Dark Matter (CDM) model of structure 
formation has been widely adopted and serves well as a framework for modeling 
the galaxy formation. The understanding of CDM halo formation is a key factor 
to understand the formation process and properties of large-scale structure 
as well as the galaxies that form and evolve within the dark halos.
There are multiple important properties regarding dark halo formation that 
have been intensively studied, including halo mass function 
~\citep[e.g.][]{BCEK91,LC93,ST99}, density profile~\citep[e.g.][]{NFW,Bul01,Lu06},
angular momentum property~\citep[e.g.][]{BE87,CL96,Bul01a},
mass accretion history~\citep[e.g.][]{Zhao03a,Wec02,Li07},
and clustering property ~\citep[e.g.][]{Mo96,Lem99,ST99,SMT01,Gao05}.

Besides the properties mentioned above, halo-halo mergers and the 
subsequent  evolution of resultant sub-halos have been of 
great interests recently, in both analytical models 
and $N$-body simulations~\citep[][]{Sheth03,Gao04a,Del04,
vdb05,CG08a,CG08b,Ang08,Wet08}. 
Since galaxies are believed to initially reside at the center of  
and merge along with dark halos, these events are therefore 
highly correlated with galaxy evolution.
In this scenario, mergers play a transitional role in converting 
central galaxies into satellite galaxies in the post-merger halos. 
They may also trigger the evolution of various galactic properties, 
such as the morphology, luminosity, color and spacial 
distribution~\citep[][]{BH96,NB03,Hop06,Maller06,Del07,Mc08}.

Although, from a phenomenological point of view, some observational 
statistics based on current sub-halo mass model matches well with the 
observations ~\citep[][]{Mand06,Kim08}, our understanding of galaxy formation 
still needs improvements due to the insufficient modeling of various 
physical processes, such as cooling, feedback, and merging history. 
It is also ambiguous how exactly post-merger galaxies are linked 
with pre-merger dark halos,  because once a merger happens,
the subsequent tidal forces and dynamical friction will cause the sub-halo, 
formerly a host halo, to loose mass and possibly become completely destroyed. 
This process gives rise to several possible fates of the stellar components 
of the galaxies that merge along with the sub-halos~\citep[][]{Yang09}.

Despite the details of how a satellite galaxy evolves in a denser environment,
it is always important to quantify the mass function of the associated sub-halos
at the time of merging, for several reasons. 
First, previous studies have suggested that mass is a key 
factor of various properties of dark halos, such as density 
profile~\citep[][]{NFW}, sub-halo population~\citep[][]{Gao04a},
clustering property~\citep[][]{Mo96}. Secondly, and more importantly 
perhaps, different approaches such as the halo occupation distribution 
(HOD) or similar models~\citep[e.g.][]{BW02,Zheng05,Tinker05,Wang06} 
and conditional luminosity function (CLF) model~\citep[][]{Yang03,vdb07},
which base their galaxy statistics on host halo mass, have resulted
in reliable descriptions of the distribution of galaxies.
Therefore, to better understand  the link  between sub-halo mergers and 
post-merger galaxies, some important issues need to be addressed.
For example, based on extended Press-Schechter formalism and direct $N$-body
simulations, ~\citet[][]{vdb05} and ~\citet[][]{CG08a} found that 
the unevolved sub-halo mass function (USMF) 
of the progenitors that merged into halo main branch follows
a universal form. Their findings are useful because the results can be linked 
to the number of central galaxies that may have turned into
satellite galaxies through {\it direct} merger into a final halo. 
However, this information is insufficient to account for {\it all} incidences
of mergers during the entire galaxy assembly history,
because the hierarchical nature of CDM model suggests that sub-halos were
independent host halos before the time of accretion, and it is likely
they inherit the generic sub-halo population by the time when they 
became sub-halos. 
Thus, to investigate the possible effects on the statistics of galaxy properties
from the angle of sub-halos, one needs to further clarify
two questions. First, is the USMF really generic (i.e., does it depend on other 
quantities such as redshift than halo mass)? Second, what may be the difference
if one takes into account the inherited sub-halo statistics of a sub-halo itself?
Today, high resolution $N$-body simulations provide a direct way to measure 
the merger statistics of dark halos to relatively high redshift with a good 
mass resolution.

In this paper, we take advantage of a large $N$-body simulation 
and its distinguished sub-halo statistics to answer the questions mentioned 
above. This paper is organized as follows. In section~\ref{sec:sim4} we give 
a brief overview of the simulation and the algorithm used to construct 
the halo merging tree.  In section~\ref{sec:USMF} we describe in detail how to 
use the merging tree to identify halo-halo mergers during the halo 
accretion history, and further derive the unevolved mass function of the 
sub-halos characterized in several different ways. In section~\ref{sec:acctime} 
we study the accretion time of sub-halos and mass function of sub-halos accreted 
at given redshift. Lastly in section~\ref{sec:concl4} we summarize our results.
\section{The Simulation}
\label{sec:sim4}
In this paper we use the  ``Millennium Simulation'' (MS) carried out by 
the Virgo Consortium~\citep{Sp05}. We have used the same simulation data in an 
earlier paper to study the age dependence of dark halo spacial 
distribution~\citep[][]{Li08}. This simulation follows the evolution
of $2160^3$ dark matter particles in a cubic box of $500\mpch$ on a side,
with a mass resolution of approximately $8.6\times 10^8\msunh$ per particle. 
The simulation 
adopts a flat $\Lambda$CDM model with $\Om=\Odm+\Obaryon=0.205+0.045=0.25$, 
where $\Odm$ and $\Obaryon$ stand for the current densities of dark matter 
and baryons respectively; the linear r.m.s. density fluctuation in a sphere 
of an $8\mpch$ radius, $\seight$, equals 0.9; and the dimensionless Hubble 
expansion parameter $h=0.73$. There are 63 snapshot outputs between $z=0$ and 
$z=80$, with a roughly even placement in $\ln(1+z)$ space. The standard 
Friends-Of-Friends (FOF) algorithm with a linking length parameter 
$b_{\rm l}=0.2$ is used to identify FOF dark halos. Only FOF halos with
more than 20 particles are resolved. Based on the FOF catalogue, each FOF 
halo is then assigned a corresponding  ``virial mass'', $\mh$, so that the 
average density contrast between the ``virial halo'' and 
cosmic critical density $\rho_{\rm c}$ is approximately 200. The
value of $\mh$ is slightly smaller (about $5-10\%$ on average) 
than the total mass of the corresponding FOF halo, and in 
general $\mh$ accounts for the mass in the central part.
This definition is therefore less affected by the so-called linking-bridge 
problem that can cause uncertainties in the halo mass.
In what follows, we always use ``virial mass'', $\mh$, as the halo mass.  
In order to ensure robustness and completeness of 
our sub-halo analysis, we only use sub-halos with masses above 
a mass limit $M_{\rm lim}= 2\times 10^{10}\msunh$.  
This mass limit is slightly higher than the re-simulated halos used 
by~\citet[][]{CG08a}, but the simulation volume of the MS allows us 
to use many more halos to gain better statistics.

The halo merging trees in the MS are constructed on the basis of 
sub-halos. In each FOF group, self-bound sub-structures (sub-halos) are further 
identified using {\sc SUBFIND}~\citep{Sp01}, with the largest ``sub-halo'' being 
the ``main'' halo. A sub-halo 1 at redshift $z_1$ is considered a progenitor of 
another sub-halo 2 at $z_2$ ($z_1 > z_2$) if the majority of its most bound 
particles are in sub-halo 2. In the literature, merging history trees based 
directly on FOF halos are widely employed to study the mass accretion history 
of dark halos. The ``sub-halo''-based linking algorithm used in the MS, 
however, has special advantages over the FOF merging tree in the  
study of the evolution of sub-halos. By definition, this algorithm 
enables a more clear-cut history tracer of sub-halos~\citep[see e.g.,][]{FM08}.

\section{Unevolved Sub-halo Mass Functions}
\label{sec:USMF}

Mergers are important events during the lifetime of a galaxy. If there exists 
a one-to-one correspondence between a central galaxy and a host halo, then all 
galaxies are initially central galaxies at some high redshift in the
hierarchical scenario of structure formation. Subsequent halo mergers play 
an crucial role in galaxy evolution, in the sense that a central galaxy will
be formally transferred into a satellite galaxy and perhaps evolve 
passively afterwards without a significant amount of star formation. 
As mentioned before, galaxy properties may be highly correlated with their 
host halo mass. Understanding the mass function of the progenitors 
of the sub-halos at the time of accretion is a key step in understanding 
the formation and evolution of satellite galaxies. In what follows 
we will refer the mass function of sub-halos at accretion as the    
the unevolved sub-halo mass function (USMF), which reflects the fact 
that the sub-halos at the times of accretion have not yet been processed 
by dynamical effects, such as tidal stripping. In the rest of this section, 
we will discuss the USMF of sub-halos in the following three categories:
\begin{enumerate}
\item In Sub-section \ref{sec:USMF_MB}, we focus on sub-halos on the main 
branch of the merging tree, i.e., progenitors that directly merge with the 
main progenitors of dark halos. The same case has been studied by~\citet[][]{CG08a}.
\item In Sub-section \ref{sec:USMF_all}, we include all sub-halos that have 
merged into the entire merging tree of a dark halo.
\item Finally in Sub-section \ref{sec:survived}, we focus on sub-halos 
that are directly identifiable in the present-day halos, the so-called 
``survived sub-halos''.
\end{enumerate}

\subsection{Main branch sub-halos}
\label{sec:USMF_MB}

\begin{figure}
\centerline{\psfig{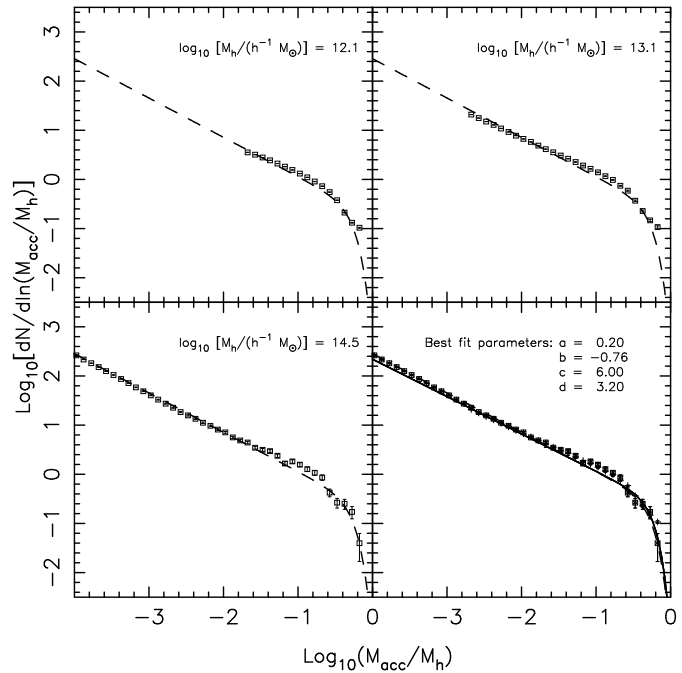}}
\caption{The USMF of main branch sub-halos. The {\it upper two panels 
and the lower left panel} show the USMF of $z_{\rm h}=0$ halos with  
$\mh = 10^{12.1}, 10^{13.1}, 10^{14.5}\msunh$, respectively. Data points
are the average over all halos with mass $\mh$, error bars represent
the standard error of the average. For reference, in each panel we also plot, 
with identical dashed lines, the best-fit USMF from~\citet[][]{CG08a}. 
In the {\it lower right panel}, 
we summarize all the data from previous panels, and plot 
equation~(\ref{eqn:usmf}) (in thick solid line) with an empirical set of 
parameters (as indicated in the panel) which provides a universal fit to 
all of our data.
}
\label{usmf_mb}
\end{figure}

To construct the halo main branch, we start with the final halo at a 
given redshift $z_{\rm h}$ (in this paper, $z_{\rm h}=0$ unless otherwise mentioned), 
and trace its most massive progenitor (the main progenitor) in the 
adjacent snapshot at higher redshift. We then repeat this procedure for 
the main progenitor till the progenitor mass is too small to be resolved. 
During this procedure, we also search the indices of all other progenitors 
that have directly merged into the main progenitor. If a progenitor was an 
independent halo before merger, we register its mass as well as the redshift 
at which it was accreted (the redshift information will be used later 
to study the mass function of sub-halos at given accretion time). 
This method eliminates cases where progenitors were already sub-halos of 
other more massive progenitors at the time of merging. With the information 
collected in this way, we are able to construct the USMF of main branch 
sub-halos. The results are plotted in Fig.~\ref{usmf_mb} for
host halos of different masses (as indicated). 

We adopt the same functional form proposed by ~\citet[][]{CG08a}
to fit the simulation results. Given a final halo (host halo) mass $\mh$ 
and a sub-halo mass {\it at accretion}, $\macc$, the USMF, $F$, is 
written as  
\begin{eqnarray}
\label{eqn:usmf}
F\left( \frac{\macc}{\mh}\right) 
 & = & \frac{{\dd}N}{{\dd} \ln (\macc/\mh)} \nonumber \\
 & = &  a\left(\frac{\macc}{\mh}\right)^{b}
    \exp\left[ -c \left(\frac{\macc}{\mh}\right)^{d}\right],
\end{eqnarray}
where $N$ stands for the number of sub-halos that were accreted 
and $a$, $b$, $c$, $d$ are fitting parameters. 
At the low-mass end ($\macc/\mh\rightarrow 0$), this is 
a power-law, while at high-mass end ($\macc/\mh\rightarrow 1$), 
the function decreases exponentially with $(\macc/\mh)^d$.
If the other parameters are fixed, $a$ represents the overall amplitude, 
$b$ indicates the low-mass end power-index, $c$ indicates the transitional 
point where the curve changes its shape, and $d$ determines the 
steepness of the exponential decline. However, different combinations of 
parameters can result in $F$ with similar shapes within the mass range 
probed here ($\log_{10}[\macc/\mh] \in [-4,0]$). Therefore we do not 
intend to fit the result for each host-halo mass bin separately.   
Instead, we use all the mass bins to obtain an overall fit, which is 
shown as the solid line in the last panel. For comparison the best-fit USMF 
obtained from~\citet[][]{CG08a} is shown in each panel of Fig.~\ref{usmf_mb}
as the dashed curve.

Our results show an overall excellent agreement with the result by 
\citet[][]{CG08a}. The values of the fitting parameters we obtain 
are very close to what were proposed by ~\citet[][]{CG08a}, with 
only slight difference. For instance, we find the low-mass end 
power-law index $b=-0.76$, which is a slightly shallower than their 
$-0.8$. Our slope is chosen so as to reconcile the slightly higher 
``shoulder'' found in the mass range $\log_{10}(\macc/\mh) 
\in [-1.5,-0.5]$. Note that for halos with $\mh = 10^{12.1}\msunh$, 
we do not have data points that cover far enough into the 
power-law part. 

\begin{figure}
\begin{center}
\psfig{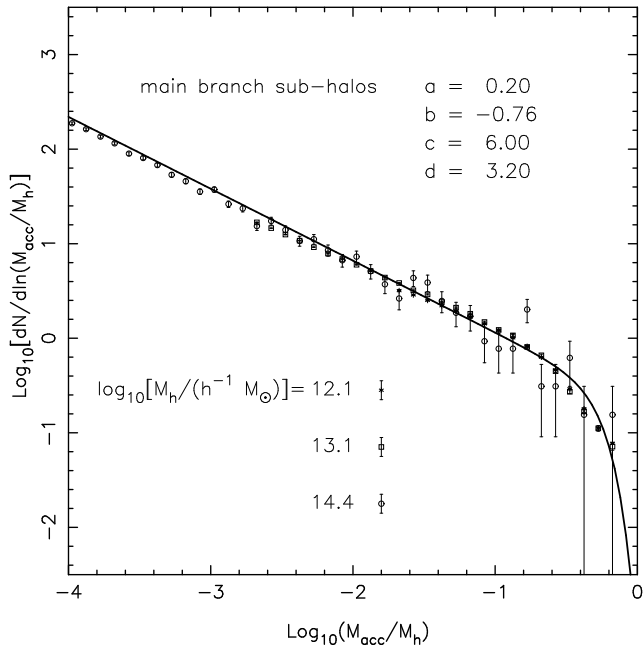}
\end{center}
\caption{The USMFs of main branch sub-halos for host halos at $z_{\rm h}=1$. 
Different symbols represent the
data points for host halos with different masses (as indicated), 
and the solid line is the universal fits we have obtained from $z_{\rm h}=0$ halos.}
\label{usmf_mb_z1}
\end{figure}

We also estimate the USMF for host halos identified at redshift $z_{\rm h}=1$,
and the result is shown in Fig.~\ref{usmf_mb_z1}. Although the cosmic density 
field has evolved significantly during the time interval from $z_{\rm h}=1$ to $z_{\rm h}=0$,
the USMF at $z_{\rm h}=1$ has the same form as that for $z_{\rm h}=0$ halos.
All these suggest that the USMF of the main branch sub-halos has 
a universal form, independent of host halo mass and the redshift 
at which the host halo is identified.

\subsection{All sub-halos on the merging tree}
\label{sec:USMF_all}

The merging history of a dark halo is in general quite complex. 
At lower redshifts, after a halo has assembled its main body, mergers may 
primarily happen on the main branch. However, at higher redshift,
when a large fraction of the final halo mass was still part of
smaller progenitors, mergers that take place on the sub-branches
of the merging tree can no longer be neglected. In addition, the 
sub-halos that merge into the sub-branches may still present at the 
time when their host halos merge into larger halos. Although 
it is likely that most sub-halos that merge at high redshift  
may have already been dissolved by dynamical friction and tidal 
stripping by the time when the final halo assembles, the satellite 
galaxies that merged along with them may be more resistant
to these dynamical effects. Therefore, it is interesting to investigate 
the statistical properties of these merging events.

In order to quantify the USMF of all sub-halos in the entire merging 
tree of a halo, we start from the final host halo and trace back to all 
its progenitors that have ever merged as a sub-halo, regardless whether 
the merger takes place on the main branch or sub-branches. Once we found 
a merger between two independent halos, we register the mass of the 
sub-halo and the time of merger. 

\begin{figure}
\centerline{\psfig{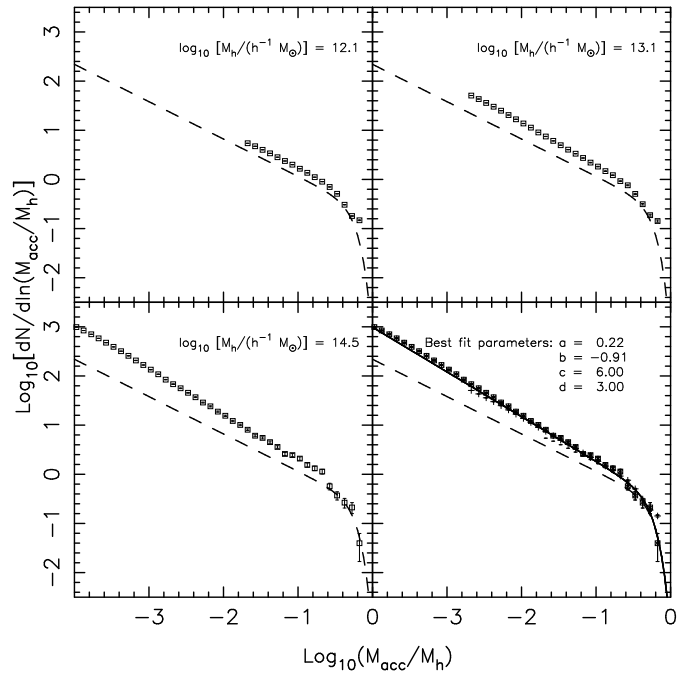}}
\caption{The USMF of all sub-halos that merged on the entire halo merging 
tree, plotted in the same way as in Fig.~\ref{usmf_mb}. 
Here, for reference purpose, the dashed lines in each panel
represent our ``universal'' fit of the USMF of the main branch sub-halos 
(the same as the thick solid line in the lower right panel of Fig.~\ref{usmf_mb}).
Similar to Fig.~\ref{usmf_mb}, in the {\it lower right panel} we choose
an empirical set of parameters (values as indicated in the panel) for 
equation~(\ref{eqn:usmf}) and plot in thick solid line,
so that it simultaneously fits all data points from the previous 
three panels.
}
\label{usmf_all}
\end{figure}

Fig.~\ref{usmf_all} shows the USMF of all sub-halos in the halo merging tree,
in the same way as Fig.~\ref{usmf_mb} for sub-halos in the main branch.
Interestingly, equation~(\ref{eqn:usmf}) still provides a good description
of the USMF in this case, although the fitting parameters are 
different from those for the sub-halos in the main branch
(see the solid line in the lower right panel and the 
values of the fitting parameters listed in the panel). 
Comparing the results here with those shown in Fig.~\ref{usmf_mb},  
we see that the overall amplitude here is higher, due to the fact 
that sub-halos on sub-branches  are also included. In addition, 
the increase in the amplitude is much larger for low-mass sub-halos than 
for massive ones, giving rise to a steeper power-law slope in the 
low-mass end -- compare the data points in each panel with the 
dashed curve that shows the fitting result of the USMF for 
sub-halos in the main branch. This is not difficult to understand. 
When we trace back in time to all branches on the merging tree, 
the number of sub-branches on the halo merging tree increases 
significantly with redshift due to bifurcation. 
Meanwhile, the average mass of progenitors drops dramatically because of 
mass conservation. Since our mass function is based on the unevolved 
merger progenitors, more mergers of low-mass sub-halos are expected 
at higher redshift. 

\begin{figure}
\begin{center}
\psfig{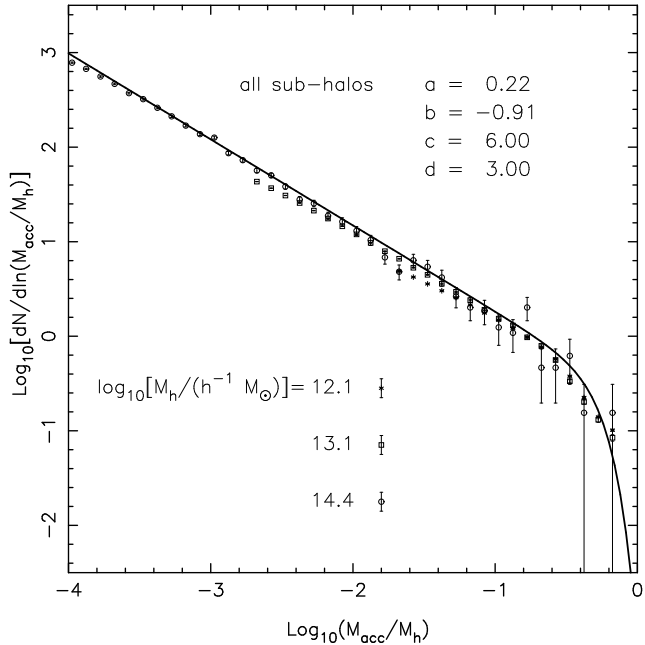}
\end{center}
\caption{The USMFs of all sub-halos for host halos at $z_{\rm h}=1$.
Same as in Fig.~\ref{usmf_mb_z1}, different symbols represent the
data points for host halos with different masses (as indicated), and the solid line
is the universal fit we have obtained from $z_{\rm h}=0$ halos.}
\label{usmf_all_z1}
\end{figure}

\begin{figure}
\centerline{\psfig{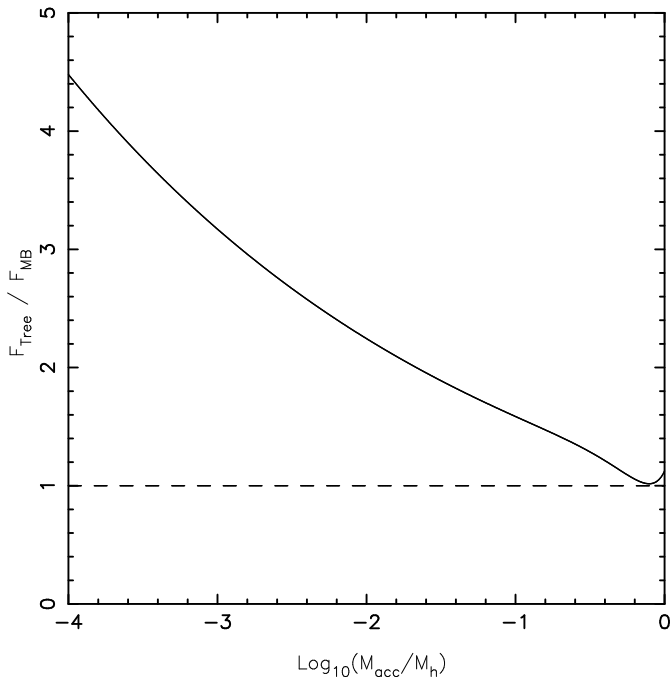}}
\caption{The solid line shows the USMF of all sub-halos
divided by the USMF of main branch sub-halos, based on the two 
fitting results we have obtained. Dashed line is a reference line of $y=1$.}
\label{usmf_diff}
\end{figure}

In Fig.~\ref{usmf_all_z1} we show the USMF of all sub-halos 
for host halos identified at $z_{\rm h}=1$. Clearly, this USMF
shows remarkable agreement with that for $z_{\rm h}=0$ host halos, 
indicating that the USMF of sub-halos in the entire merging tree also has a 
universal form. 

Let $F_{\rm Tree}$ and $F_{\rm MB}$ represent our universal fits to the 
USMFs of all sub-halos and the main branch sub-halos, respectively.
Fig.~\ref{usmf_diff} shows the ratio of these two functions 
as a function of $\macc/\mh$. At the low-mass end $F_{\rm Tree}$ is about 
four times $F_{\rm MB}$, while at the high-mass end they are nearly equal. 
The significant excessive rate of mergers at low-mass end seen in the 
ratio indicates the abundance of sub-halos that were accreted by the 
sub-branches of the merging tree (we will discuss in details later).
These sub-halos may  end up as the so-called sub-subhalos when they 
finally settle in the main progenitor \citep[][]{Yang09}.

\begin{figure}
\centerline{\psfig{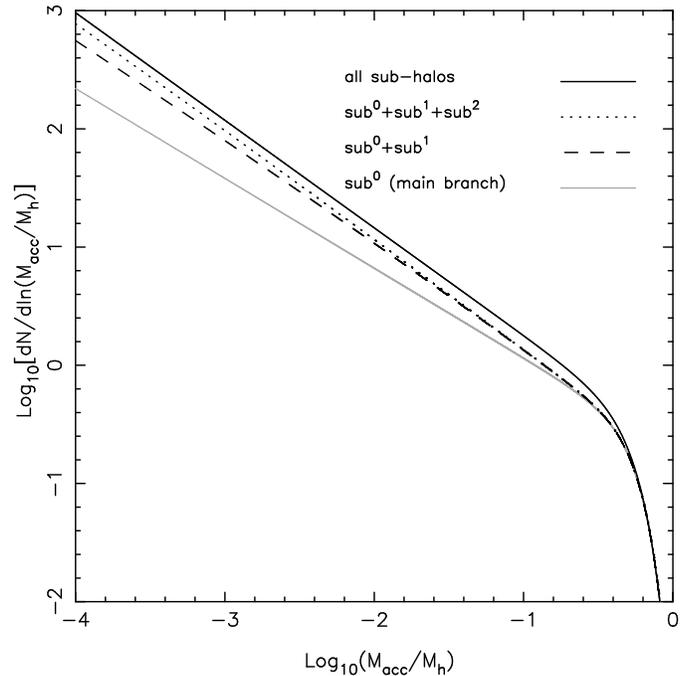}}
\caption{Comparison between the USMF of main branch sub-halos (gray solid line,
which is same as the solid line in the lower right panel of Fig.~\ref{usmf_mb}),
all sub-halos (dark solid line, the same as the solid line
in the lower right panel of Fig.~\ref{usmf_all}), the sum of sub$^0$ and 
sub$^1$-halos (dashed line), as well as the sum of sub$^0$, sub$^1$ and 
sub$^2$ -halos (short-dashed line) from the model prediction 
(equation~[\ref{eqn:usmf3}]).}
\label{subi}
\end{figure}

 As we have seen, the USMF of the main branch sub-halos is 
universal, independent of the mass and redshift of host halos.
This proposition has been adopted by some authors when modeling the 
population of satellite galaxies in dark matter halos \citep[e.g.][]{Yang09}. 
\citet[][]{Yang09} assumed that the USMF of main branch sub-halos is 
self-similar, and sub-halos can be divided into different ``levels''. 
Since sub-halos can themselves be considered as host halos
at the time of accretion, their sub-halos (referred to sub-sub-halos, 
or sub$^1$-halos) are also expected to obey the universal USMF. Similarly, 
all levels of sub-halos (sub$^i$-halos, $i=0, 1, 2, 3, \cdot\cdot\cdot$, 
where superscript `$0$' stands for the main branch sub-halos) 
should have the same form of USMF. The summation of the USMFs at 
all levels should be equal to the USMF of sub-halos in the whole tree.  
To test this, we rewrite equation~(\ref{eqn:usmf}) as
\begin{eqnarray}
\label{eqn:usmf2}
n_{\rm un,0}(\macc|\mh)& = & \frac{{\dd}N}{{\dd} \macc}  \\
& = &  \frac{a}{\mh} \left(\frac{\macc}{\mh}\right)^{b-1}
\exp\left[ -c \left(\frac{\macc}{\mh}\right)^{d}\right].
\nonumber
\end{eqnarray}
Since equation~(\ref{eqn:usmf2}) is universal, it should apply to all 
sub$^i$-halos ($i=0,1,2,3,\cdot\cdot\cdot$). This allows us to calculate the 
conditional USMF of sub$^i$-halos given the host halo mass $\mh$,
\begin{eqnarray}
\label{eqn:usmf3}
\lefteqn{n_{\rm un,i}(M_{{\rm acc},i}|\mh)=} \nonumber\\
&&\int^{\mh}_{0} n_{\rm un,0}(M_{{\rm acc},i}|\macc)
n_{\rm un,i-1}(\macc|\mh)\dd \macc . 
\end{eqnarray}
Fig.~\ref{subi} shows the comparison between the USMF of main branch 
sub-halos, all sub-halos, the sum of sub$^0$ and sub$^1$-halos, 
as well as the sum of sub$^0$, sub$^1$ and sub$^2$ -halos predicted by 
equation~(\ref{eqn:usmf3}). The best-fit parameters used in the calculation 
are indicated in the lower-right panels of Figs.~\ref{usmf_mb} 
and \ref{usmf_all}, respectively. Clearly, as we include more levels of 
sub-halos, the summation of their USMF approaches asymptotically 
to that of all sub-halos. It is also interesting that sub$^1$-halos
contribute the largest fraction of small sub-halos that are not included 
in the USMF of the main branch sub-halos. Note that in the mass range 
$\macc/\mh \in [-2,-0.3]$, the difference between the dark solid line and 
the short-dashed line in Fig.~\ref{subi} is more significant. 
It is unclear if this difference is real, or it is due to the 
limited statistics of the simulation data.

\subsection{Survived sub-halos}
\label{sec:survived}

In the two cases discussed above, the USMFs do not seem to depend on 
the final halo mass or redshift, and appear to be ``universal''.
However, once a sub-halo merges into a host halo, it will undergo 
a number of non-linear processes such as dynamical friction, which 
causes the sub-halo to merge into the center of the host,   
and tidal stripping, which causes it to lose mass or to be 
completely destroyed. Therefore, the number of {\it survived} sub-halos 
may be significantly lower than the sub-halo abundance described by the 
USMF. Note that there are two kinds of survived sub-halos: those that 
directly merged into the main branch, and those that were already a 
sub-halo of a larger progenitor when being accreted by the main 
progenitor. Throughout this paper, we refer to the former as 
sub$^A$-halos, and the latter as sub$^B$-halos, which are also 
known as sub-subhalos.

\begin{figure}
\psfig{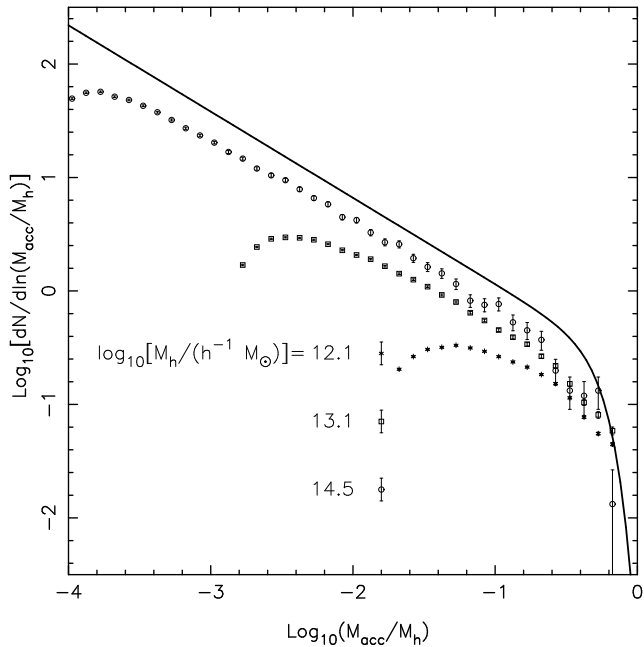}
\caption{The ``unevolved'' mass function of sub$^A$-halos, 
$F^\prime_{\rm sub^A}$. Different symbols represent different final host 
halo mass. The thick solid line is the universal form of the USMF 
of main branch sub-halos, $F_{\rm MB}$.}
\label{usmf_subs}
\end{figure}

After removing the sub$^B$-halos from our survived sub-halo catalogue, 
we construct  the ``unevolved'' mass function of the  
sub$^A$-halos. The quotation marks are used to indicate that 
a certain fraction of the main branch sub-halos have been completely 
destroyed, although the sub-halo mass used here is the mass at the time 
of accretion, $\macc$. Since the destroyed sub-halos are not included, 
we use $F^\prime_{\rm sub^A}$ to distinguish this ``unevolved''
sub-halo mass function from the USMF discussed previously.
Fig.~\ref{usmf_subs} shows the ``unevolved'' sub-halo mass function 
so defined for host halos with $\mh = 10^{12.1}$, 
$10^{13.1}$, and $10^{14.5}\msunh$, respectively. Apparently the shape 
of $F^\prime_{\rm sub^A}$ depends strongly on host halo mass. 
Unlike the USMFs discussed previously, for given host halo mass $\mh$,
$F^\prime_{\rm sub^A}$ is not a monotonic decreasing function of 
sub-halo mass, but rather, its amplitude lowers when the sub-halo 
mass becomes very small. This is caused by the dynamical 
processes after the accretion of sub-halo. Note, however, 
that the value of $\macc/\mh$ at which $F^\prime_{\rm sub^A}$
peaks depends on the mass limit $M_{\rm lim}$ adopted.
There is a high probability that a sub-halo initially 
accreted with mass $\macc$ slightly above $M_{\rm lim}$ to become 
smaller than $M_{\rm lim}$ during the post-accretion phase and thus 
to be marked as ``destroyed''. In addition, smaller sub-halos are 
more difficult to survive, because on average they were 
accreted into their hosts earlier.

Fig.~\ref{usmf_subs} may be used to estimate the number fraction
of sub-halos that survive the mass-loss process. For example, for halos 
with $\mh\sim 10^{14.5}\msunh$, about $62\%$ of the accreted main branch 
sub-halos above the mass limit $M_{\rm lim}$ have been completely 
destroyed, this fraction increases to $\sim 78\%$ and $\sim 84\%$ 
for $\mh=10^{13}\msunh$ and $\mh=10^{12}\msunh$ halos, respectively. 
This trend may be understood since small systems start to accrete progenitors 
earlier, and so their main branch sub-halos are subject to mass loss and 
destruction for a longer time. The shape of $F^\prime_{\rm sub^A}$ for 
host halos at $z_{\rm h}=1$ is similar to that at $z_{\rm h}=0$. However, the similarity 
here is less meaningful, because the shape of $F^\prime$ is highly 
affected by the non-linear effects during sub-halo mergers, which is 
a very stochastic process~\citep[][]{Ang08}.

\begin{figure}
\psfig{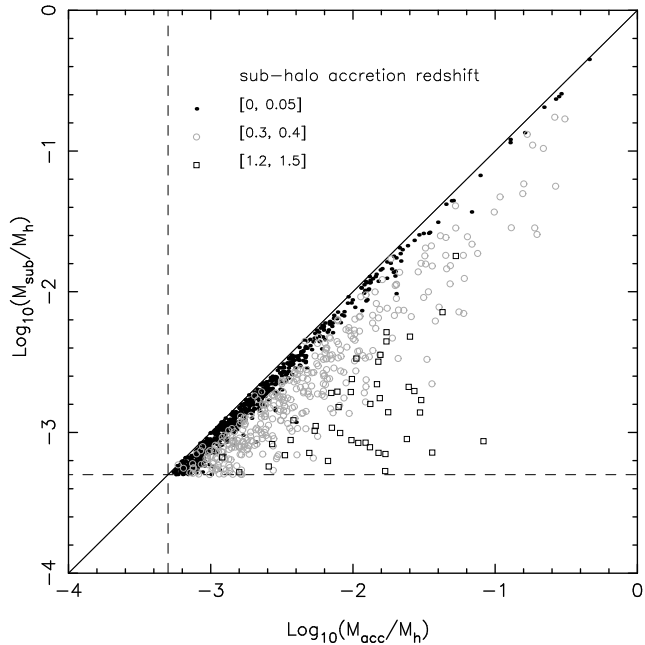}
\caption{$\macc$ against $M_{\rm sub}$ for host halos of a given mass 
$\mh$. Points are from 200 randomly selected host halos with 
$\mh \approx10^{13.6}\msunh$. Different symbols denote different redshift 
intervals during which the sub-halos enter the main progenitor of the 
host halo. Dashed lines indicate the mass limitation in our analysis, 
$M_{\rm lim}$.}
\label{macc_msub}
\end{figure}

The result presented here is consistent with that of 
\citet[][their Fig.~4]{CG08a}, although their result is based on the 
{\it evolved} sub-halo mass function. 
They found that for small host halos, there are less sub-halos
with the same fractional mass, $M_{\rm sub}/\mh$ (where $M_{\rm sub}$ is
the {\it current} mass of survived sub-halos), than more massive host halos.
In Fig.~\ref{usmf_subs}, we have showed that for smaller host halo mass,
the amplitude of the ``unevolved'' sub$^A$-halo mass function, 
$F^\prime_{\rm sub^A}$, is also lower. In addition, according 
to~\citet[][and reference therein]{CG08a}, for evolved sub-halos mass 
function, the low-mass end is always higher than the high-mass end. 
Combining their results with our results of $F^\prime_{\rm sub^A}$, 
we see that the majority of the smallest survived sub-halos 
are not the descendants of the smallest sub-halos initially accreted, 
but rather, the descendants of those that are several times more 
massive. Fig.~\ref{macc_msub} plots $\macc$, the mass at accretion, 
against the current mass, $M_{\rm sub}$ of sub-halos, in 200 host 
halos with $\mh \approx10^{13.6}\msunh$. Three different symbols denote 
different sub-halo accretion redshifts. It is clear that given 
$M_{\rm sub}$, sub-halos accreted earlier generally have higher $\macc$.
At very low redshift ($z\in [0,0.05]$), sub-halos with a wide range 
of mass ($\log_{10}[\macc/\mh] \in [-3.3,-0.5]$) have 
been accreted by the main progenitor, and they have barely suffered from 
the mass loss so that $M_{\rm sub} \approx \macc$.
However, for sub-halos that were accreted at high redshift 
($z\in[1.2,1.5]$), their $\macc$ are in general several times higher 
than $M_{\rm sub}$. 

\begin{figure}
\psfig{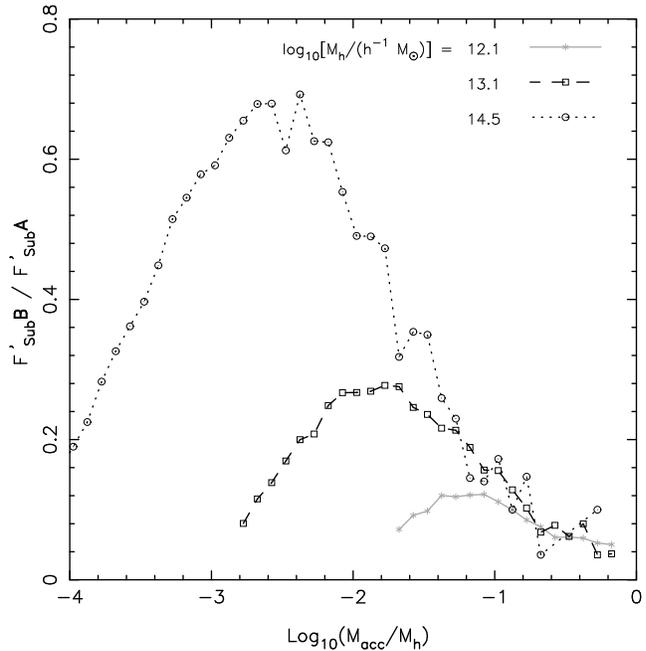}
\caption{The ratio of $F^\prime_{\rm sub^B}$
to $F^\prime_{\rm sub^A}$ (see text for details).
Different symbols indicate different host halo mass.}
\label{subhalo_diff}
\end{figure}

Besides $F^\prime_{\rm sub^A}$, we also construct the ``unevolved'' mass function
for the sub$^B$-halo population, $F^\prime_{\rm sub^B}$, in the same way
as $F^\prime_{\rm sub^A}$. Fig.~\ref{subhalo_diff} shows the ratio of 
$F^\prime_{\rm sub^B}$ to $F^\prime_{\rm sub^A}$, for host halos with 
the same masses as in Fig.~\ref{usmf_subs}. We would like to remind the
reader, once again, that the sub-halo mass used here is measured at the
time when they were last found as isolated halos. Given a sub-halo mass, 
the vertical axis in Fig.~\ref{subhalo_diff} is the ratio of the number 
of survived sub-halos initially accreted by sub-branches to the number 
of survived sub-halos initially accreted by the main branch. 
In general, $F^\prime_{\rm sub^B}/F^\prime_{\rm sub^A}$ is higher 
for massive host halos. For a given host halo mass, though, this ratio is 
always low ($\sim 0.05$) at the high-mass end ($\log_{10}[\macc/\mh] > -0.7$), 
because mergers involving sub-halos with mass comparable to that of the 
final host halo can only happen on the main progenitor at very late time.
There also appears to be a generally increasing trend in this ratio as 
sub-halo mass decreases down to a certain point.  This may be due to
two reasons. First, some small sub-halos that merge to sub-branches
of the merging tree may survive if the time scale for disruption is 
long. Second, as the redshift increases, the number of mergers that happen on 
sub-branches is not negligible. The increasing trend changes its sign 
when sub-halo mass becomes very small. The reason is that small 
sub-halos that are able to survive were most likely accreted
in the recent past, when main branch already dominates the merger 
incidences.

 We can estimate the number fraction of sub$^B$-halos among the whole 
survived sub-halo population, based on Fig.~\ref{subhalo_diff}. This 
fraction is $9\%$, $17\%$ and $28\%$, for host halo with $\mh = 10^{12.1}$, 
$10^{13.1}$, and $10^{14.5}\msunh$, respectively. Clearly, a significant 
fraction of sub-structures were sub-subhalos.

\section{Accretion time of sub-halos}
\label{sec:acctime}

Although the USMFs give a quantitative description on the abundance of
accreted sub-halos in the halo assembly history, it does not include 
the time (redshift) when the accretion happens. In galaxy formation models, 
the epoch when central galaxies became satellites is crucial as
the physical processes relevant to galaxy evolution 
after the merger are expected to be different. It is therefore 
important to incorporate the sub-halo abundance at different redshift 
into our analysis.

\subsection{Sub-halo mass function at given accretion time}

\subsubsection{Main branch sub-halos and all sub-halos}

We define the mass function of sub-halos at given accretion time 
(redshift) as follows,
\begin{equation}
\label{eqn:usmfdz}
f(z)=\frac{\dd F}{\dd z} = \frac{{\dd}N(z)}{{\dd} \ln (\macc/\mh) \dd z}, 
\label{eqn:diff}
\end{equation}
where $F$ is the USMF, $\macc$ and $\mh$ stand for the mass of sub-halos 
at the time of accretion and the mass of final host halo, respectively.
To obtain $f(z)$, we choose a redshift interval $\Delta z$ around a 
given $z$, and only count the number, $N(z)$, of sub-halos accreted 
during $\Delta z$. Within the redshift range of interste, 
we found $\Delta z\sim 0.1$ effectively eliminates 
the noise and result in a relatively smooth shape of $f$.

\begin{figure}
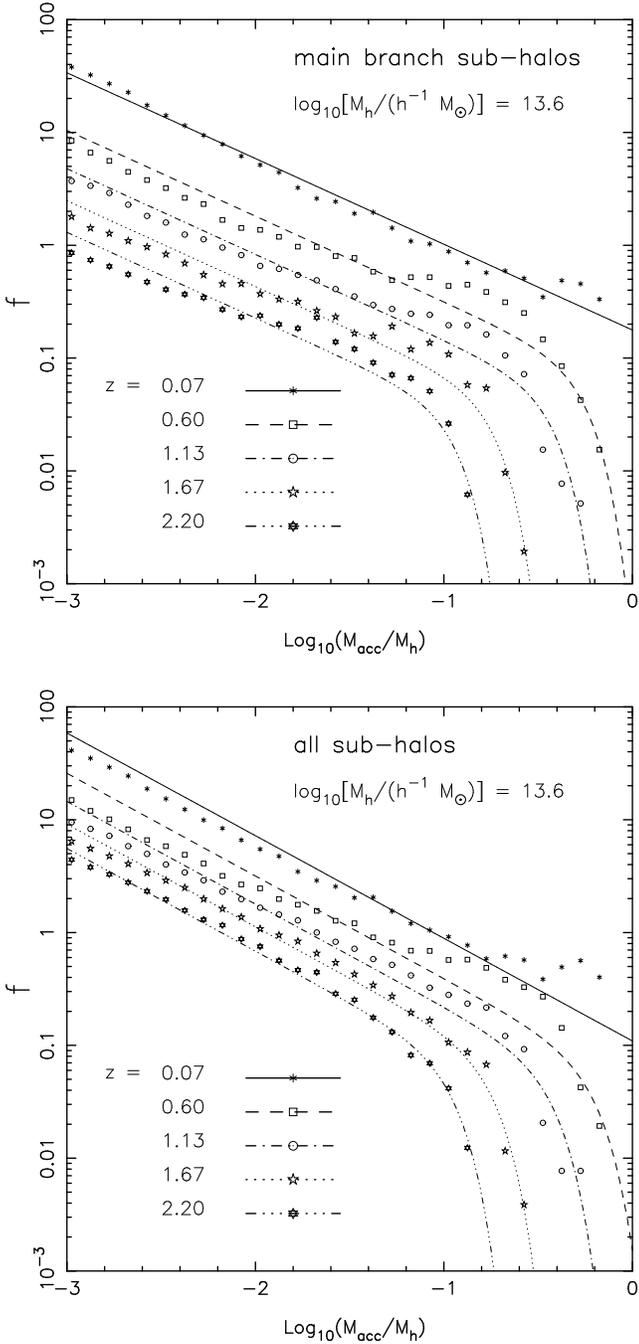

\begin{center}
\vbox{
\psfig{file=fig10a.ps,angle=270,width=1.0\hsize}
\bigskip
\psfig{file=fig10b.ps,angle=270,width=1.0\hsize}
}
\end{center}
\caption{The mass function at accretion, $f$, of main branch (upper panel) 
and all (lower panel) sub-halos, given accretion redshift $z$ and
host halo mass $10^{13.6}\msunh$.
Different symbols and lines represent the data points and their best fits
according to equation~(\ref{eqn:usmf})
(with fixed $b$ and $d$, see text for details), at different redshifts. 
}
\label{usmf_zdir}
\end{figure}

Fig.~\ref{usmf_zdir} shows $f$ of main branch sub-halos 
and all sub-halos for host halos with $\mh = 10^{13.6}\msunh$. 
Interestingly, in each case, $f$ can still be described by 
equation~(\ref{eqn:usmf}) reasonably well. In addition, we found that 
the low-mass end  power-index $b$ of $f$ are virtually independent of $z$, 
and is quite similar to the power-index we have obtained from the 
corresponding USMF. Since $F=\int f\dd z$, it is expected that the 
integration of $f$ over $z$ reproduces the low-mass end power-index of $F$.
The exponential shape of $f$ (described by $d$) at the high 
mass end also shows no obvious dependence on $z$. On the other hand, 
the amplitude of $f$ and the transitional point where $f$ deviates 
from the power-law clearly depend on the redshift. 
By keeping $b$ and $d$ fixed at the values obtained from the USMFs 
($b=-0.76$, $d=3.2$ for main branch sub-halos, and 
$b=-0.91$, $d=3.0$ for all sub-halos), we fit $f$ according to 
equation~(\ref{eqn:usmf}). Styled lines in Fig.~\ref{usmf_zdir}
are the best-fits of $f$ so obtained at the corresponding redshift. 

\begin{figure*}
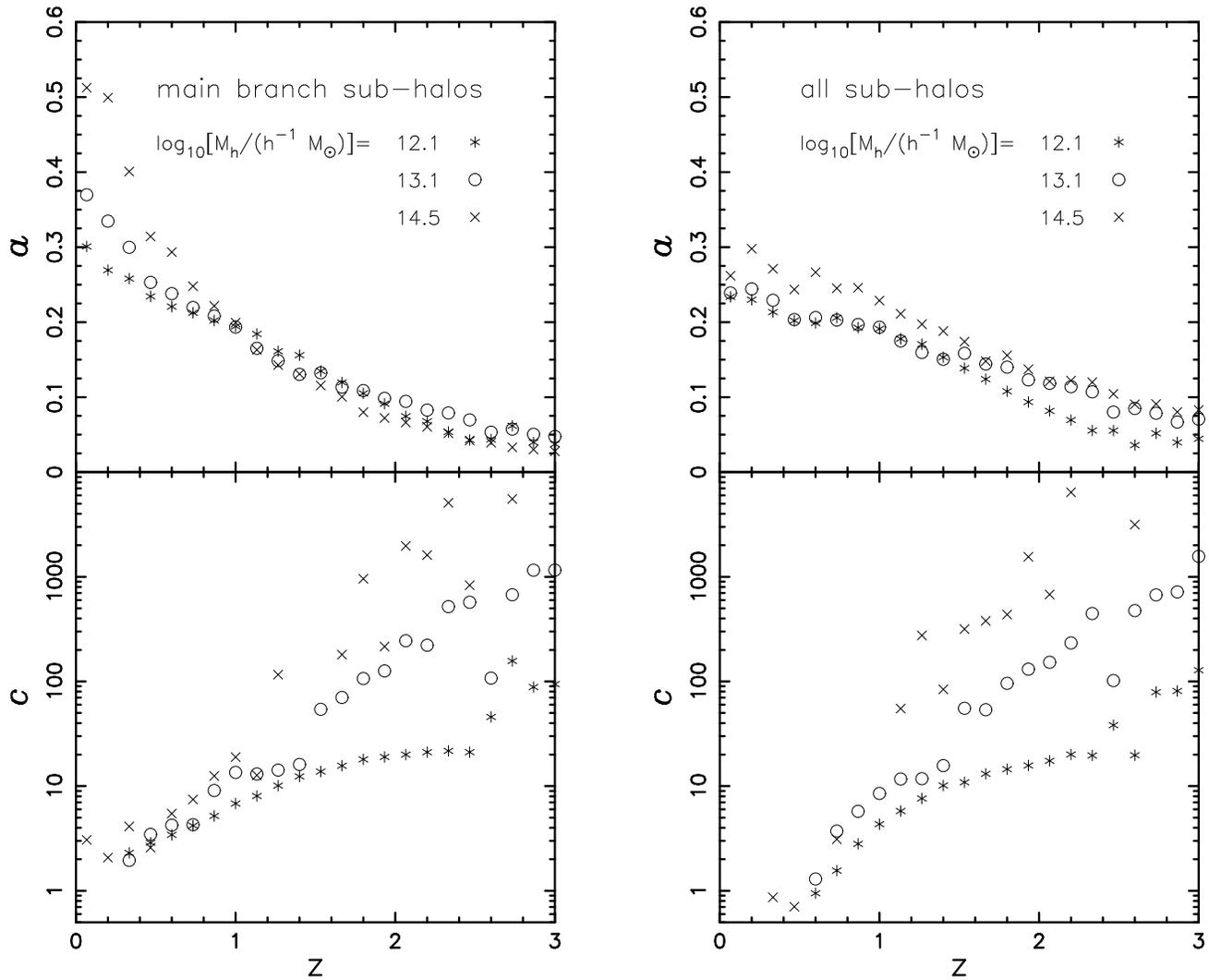

\begin{center}
\hbox{
\psfig{file=fig11a.ps,angle=270,width=0.45\hsize}
\qquad \qquad
\psfig{file=fig11b.ps,angle=270,width=0.45\hsize}
}
\end{center}
\caption{Best fit parameters $a$ and $c$, given fixed $b$ and $d$, of $f$,
 against redshift $z$.  {\it Panels on the left} shows the result for 
main branch sub-halos, {\it panels on the right} shows the result for all 
sub-halos. Different symbols represent different host halo masses, 
as indicated in the figure.
}
\label{acfit}
\end{figure*}

In Fig.~\ref{acfit}, we show the best-fit $a$ and $c$ against
the redshift $z$, for host halo with different masses. Panels on the left
are best-fit $a$ and $c$ for main branch sub-halos, while panels on the 
right are best-fit $a$ and $c$ for all sub-halos.  In general, $a$ 
always decreases monotonically as $z$ increases,
which implies that more sub-halos are accreted at lower redshift,
especially for massive halos. Meanwhile, $c$ shows positive correlation 
with $z$, which means that, compared with small sub-halos, 
the number of massive sub-halos drops more quickly as redshift
increases. This disagrees with the result of~\citet[][]{CG08a}.
Their Fig. 1 shows that the USMF of the main branch sub-halos accreted 
before the halo formation time $z_{\rm f}$ is identical to the USMF 
of the sub-halos accreted after $z_{\rm f}$, with proper adjustment 
in the amplitude $a$ only. However, as we just mentioned, the number 
of massive sub-halos drops more quickly at higher redshift, 
and therefore simply offsetting the USMF of high-redshift sub-halos 
along the vertical direction cannot reconcile the lack of massive 
sub-halos and reproduce the shape at high-mass end of the USMF.
Our results suggest that the relative abundance of massive sub-halos 
becomes higher at low redshift, consistent with the hierarchical
formation of dark halos in  a CDM model.

\subsubsection{Survived sub-halos accreted by the main progenitor}

Given the time of merging, let us look at the mass function of 
sub-halos that survive as sub-structures in the final halo. 
We focus on sub$^A$-halos, as the results for sub$^B$-halos are 
similar. Based on the sub$^A$-halo catalogue, we can register
the time when they first became satellites of the main 
progenitor. We use $f^\prime_{\rm sub^A}(z)$ to indicate
the same sub$^A$-halo mass function at given accretion time
defined in equation~(\ref{eqn:diff}).

\begin{figure}
\psfig{file=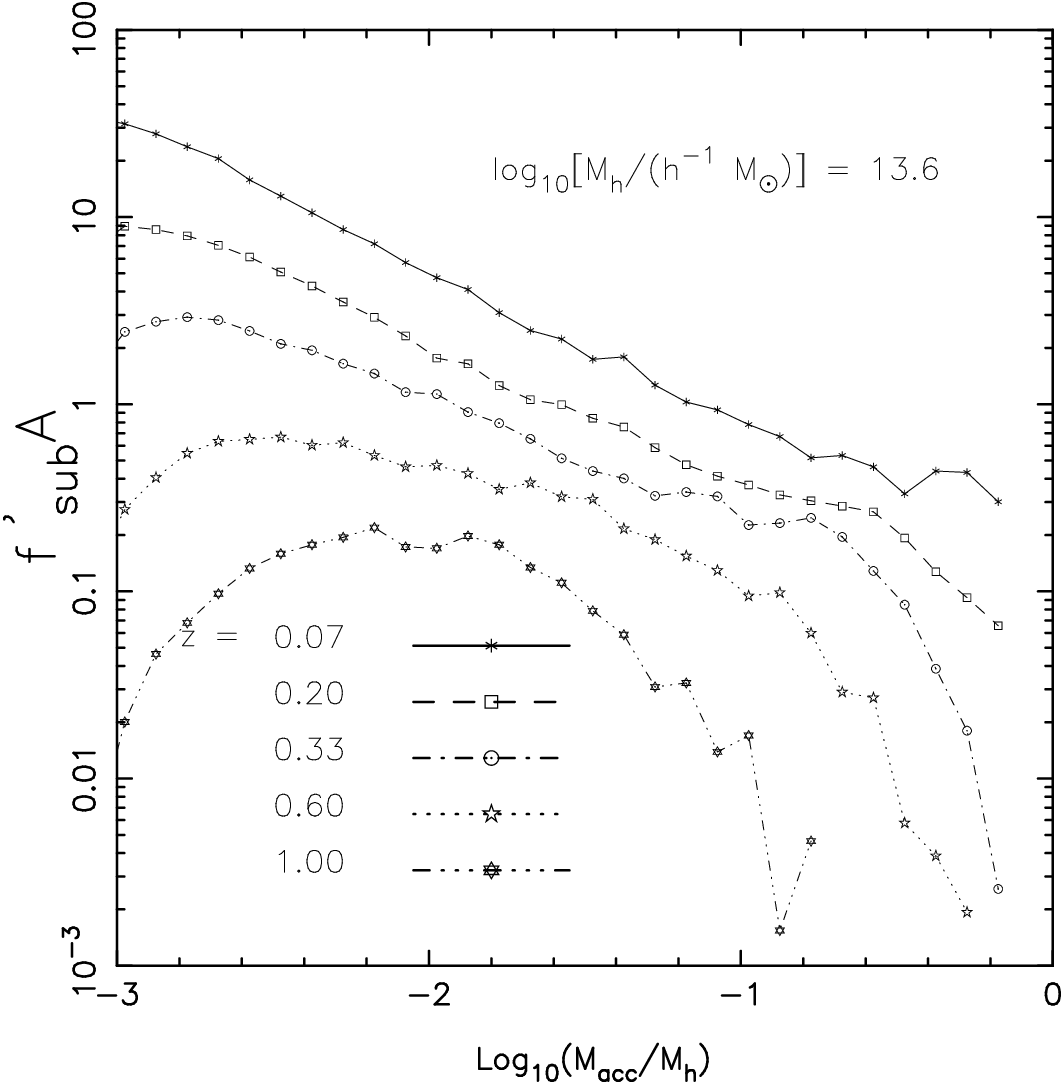,angle=270,width=1.0\hsize}
\caption{
The sub$^A$-halo mass function at accretion, $f^\prime_{\rm sub^A}$, 
given accretion redshift $z$ and host halo mass $\mh=10^{13.6}\msunh$.
Different symbols connected with styled lines represent results 
at different redshifts.
}
\label{usmf_zdir_sub0}
\end{figure}

Fig.~\ref{usmf_zdir_sub0} shows $f^\prime_{\rm sub^A}$ at
different redshifts, for host halo mass $\mh =10^{13.6}\msunh$. 
Note that the redshifts we used to plot $f^\prime_{\rm sub^A}(z)$
is, on average, lower than the redshifts used in Fig.~\ref{usmf_zdir}, 
because at higher redshift such as $z>1$, $f^\prime_{\rm sub^A}$
becomes extremely small. Comparing  Fig.~\ref{usmf_zdir_sub0}
with the {\it upper panel} of Fig.~\ref{usmf_zdir}, one can find both 
similarity and difference. At very low redshift ($z=0.07$), 
$f^\prime_{\rm sub^A}$ and $f$ are similar, due to the fact that 
sub-halos accreted by the main progenitor recently have a high 
survival rate. However, at higher redshift ($z=0.6$),  
$f^\prime_{\rm sub^A}$ becomes much lower than $f$, owing to the 
dynamical effects that can effectively destroy the sub-halos accreted 
at early time.

As we have shown in Fig.~\ref{usmf_subs}, the ``unevolved'' mass 
function of sub$^A$-halos, $F^\prime_{\rm sub^A}$, is not universal. 
Besides, the overall amplitude of $F^\prime_{\rm sub^A}$, also 
deviates substantially from the original USMF of main branch sub-halos, 
$F_{\rm MB}$, especially at the low-mass end.
The reason is clearly demonstrated in Fig.~\ref{usmf_zdir_sub0}.
When redshift increases, $f^\prime_{\rm sub^A}$ becomes increasingly 
lower, especially for small sub-halos. Since $F^\prime$ is the 
integration of $f^\prime$ over $z$, it is therefore expected that 
$F^\prime_{\rm sub^A}$ would have the behavior shown in Fig.~\ref{usmf_subs}.
 
\subsection{Distribution of sub-halo accretion time}

In the previous sub-section, we have discussed the sub-halo mass 
function at accretion for given redshift. It clearly shows
that the abundance of sub-halo accretion varies with 
redshift. In general, more sub-halos were accreted at lower 
redshift. It also seems that sub-halos with different masses may 
be accreted at different time. 

\begin{figure*}
\psfig{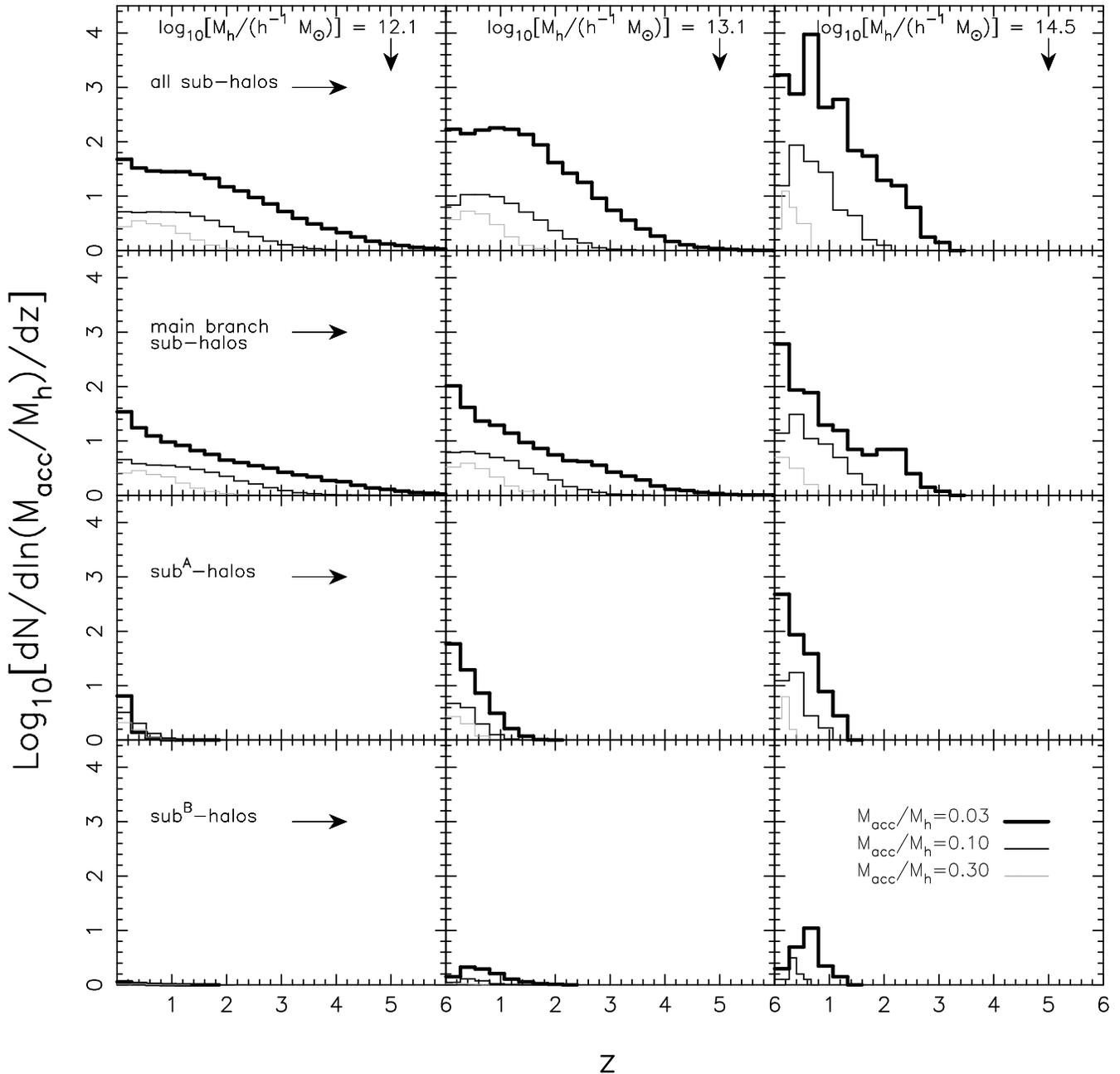}
\caption{
Sub-halo mass function at the time of accretion against redshift $z$,
given sub-halos mass $\macc$ and final host halo mass $\mh$. Each row
represents one definition of sub-halo, and different columns represent
different host halo masses, as indicated by the arrows. 
There are three lines in every panel. Think solid line is for
sub-halo with $\macc=0.03\mh$, light solid line is for sub-halo
with $\macc=0.1\mh$ and gray solid line is for sub-halo with 
$\macc=0.3\mh$.
}
\label{z_dn}
\end{figure*}

Given sub-halo mass fraction $\macc/\mh$ and host halo mass $\mh$, 
Fig.~\ref{z_dn} shows the number of sub-halo at the time of accretion 
as a function of redshift. Clearly, for fixed $\macc/\mh$, small 
systems start to accrete sub-halo earlier. For instance, dark halos 
with $10^{12.1}\msunh$ begin to acquire sub-halos with $\macc=0.03\mh$ 
at $z=5\sim 6$, while for halos with $10^{14.5}\msunh$, this happens 
at $z\sim 3$. Compared with small sub-halos, large sub-halos enter 
the system fairly late. Nearly all sub-halos with mass $\macc=0.3\mh$
enter their host at redshift $z<1.5$. 

For fixed host halo mass, large fraction of small sub-halos enter 
the system through sub-branches, especially at high redshift such 
as $z>1$, while massive sub-halos (i.e., $\macc/\mh=0.3$) enter the 
systems only through the main branch, at relatively lower redshift. 
In addition, as discussed in Section~\ref{sec:USMF}, almost all 
survived sub-halos (sub$^{A,B}$-halos) were accreted at redshift 
$z<1$, and more sub$^{A,B}$-halos are likely to survive in massive 
systems. 

\begin{figure}
\psfig{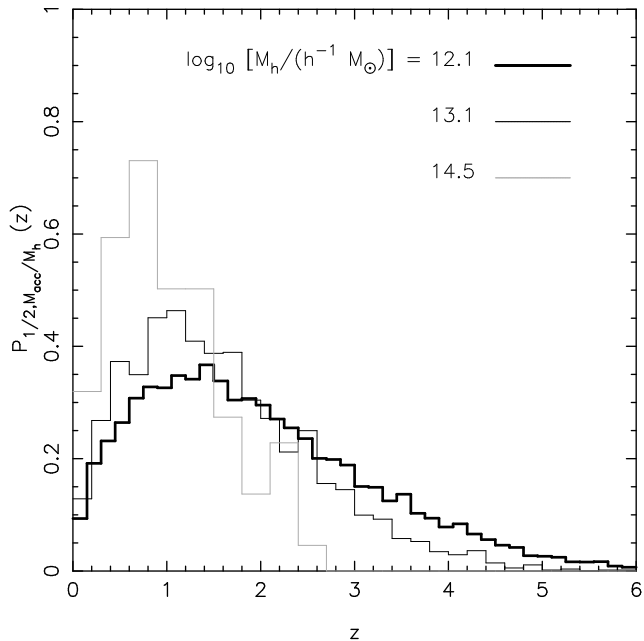}
\caption{$P_{1/2,\macc/\mh}(z)$, given main branch sub-halo mass 
$\macc = (2-5)\% \mh$. Different lines represent different host 
halo masses.}
\label{z_n05}
\end{figure}
On average, sub-halo accretion of dark halos is determined by 
the initial CDM density power-spectrum and shows hierarchical signature.
The sub-halo accretion for individual dark halos, however, can be very 
stochastic. Let $P_{1/2,\macc/\mh}(z)$ denotes the probability 
distribution function (PDF) of the redshift $z$ by which the host halo 
has acquired $1/2$ of the total number of the {\it main branch} 
sub-halos  with fixed mass $\macc/\mh$. Fig.~\ref{z_n05} shows 
$P_{1/2,\macc/\mh}$ as a function of $z$, for sub-halos with mass
$\macc = (2-5)\% \mh$. We choose $\macc \sim 3\%$ of $\mh$ to ensure that 
the majority ($>75\%$) of our catalogued halos would have more than one 
sub-halo mergers with such sub-halo mass. Clearly, the redshift covers a 
wide range: $z \in (0,6), (0,5), (0,3)$ for halos with $\mh = 10^{12.1}$, 
$10^{13.1}$, and $10^{14.5}\msunh$, respectively. This indicates that, 
even for the same sub-halo mass and host halo mass, sub-halo merger is 
a highly stochastic process.

\section{Summary}
\label{sec:concl4}

Halo-halo merger is the basis of galaxy merger.
The time of merger and the sub-halo mass at the time of merger 
are two important halo properties relevant for modeling galaxy formation.
In this paper, we study the mass function and other properties of 
sub-halo mergers during the dark halo assembly history. 
We studied three kinds of sub-halos: main branch sub-halos, all sub-halos,
and sub-halos that survived the dynamical disruption after merger. 
We also studied the redshift dependence and evolution of sub-halo 
mass function, as well as the distribution of the redshift at which 
a sub-halo is accreted. Our main findings can be summarized as follows:
\begin{enumerate}
\item We confirmed the previous result that the average unevolved 
mass function of  main branch sub-halos follows a universal functional form, 
regardless of host halo mass~\citep[][]{CG08a}. In addition, we found 
that this function is also independent of the redshift of the host halo.
\item The unevolved mass function of all sub-halos that have been accreted 
during the entire halo assembly history is also a universal function that
shows no host-halo-mass or redshift dependence. 
\item There are roughly the same or double number of sub-halos, with mass $1\%$ or $0.1\%$ of the final 
host halo mass, that were accreted by progenitors other than the 
main progenitor. The amount is significant considering the central galaxies 
that merge along with such sub-halos may be more resistant to dynamical 
disruption that destroy the sub-halos.
\item The mass function of survived sub-halos at the time of merging 
is not universal, due to the fact that large fraction of sub-halos 
that merged at early time are destroyed by dynamical friction and tidal 
stripping. The fraction of sub-subhalos can account for up to 30\% of 
the whole survived sub-halo catalogue in cluster-sized dark halos, 
and decreases with host halo mass. 
\item In general, more sub-halos are accreted at lower redshift. 
However, for given host halo and sub-halo mass, the accretion time 
has very broad distribution. Survived sub-halos are accreted 
late and therefore represent a very special subset of the total 
sub-halo population accreted into host halos.
\end{enumerate}

\section*{Acknowledgments}
The {\it Millennium Simulation} was carried out as part of the
programme of the Virgo Consortium on the Regatta supercomputer of the
Computing Centre of the Max-Planck Society in Garching.
YL would like to thank Liang Gao for the help on the simulation
data.  HJM would like to acknowledge the support of 
NSF AST-0607535, NASA AISR-126270 and NSF IIS-0611948.

\label{lastpage}


\bigskip

\begin{thebibliography}{}

\bibitem[Angulo \etal (2008)]{Ang08}
Angulo R.~E., Lacey C.~G., Baugh C.~M., Frenk C.~S., 2008, preprint (astro-ph/0810.2177)

\bibitem[Barnes \& Efstathiou (1987)]{BE87}
Barnes J., Efstathiou G., 1987, \apj, 319, 575

\bibitem[Barnes \& Hernquist (1996)]{BH96}
Barnes J., Hernquist L., 1996, \apj, 471, 115

\bibitem[Berlind \& Weinberg (2002)]{BW02}
Berlind A.~A., Weinberg D.~H., 2002, \apj, 575, 587

\bibitem[Bond \etal (1991)]{BCEK91}
Bond J.~R., Cole S., Efstathiou G., Kaiser N., 1991, \apj, 379, 440

\bibitem[Bullock \etal (2001a)]{Bul01a}
Bullock J.~S., Dekel A., Kolatt T.~S., Kravtsov A.~V., Klypin A.~A., 
Porciani C., Primack J.~R., 2001a, \apj, 555, 240

\bibitem[Bullock \etal (2001b)]{Bul01}
Bullock J.~S., Kolatt T.~S., Sigad Y., Somerville R.~S., Kravtsov A.~V.,
Klypin A.~A., Primack J.~R., Dekel A., 2001b, \mnras , 321, 559

\bibitem[Cole \& Lacey (1996)]{CL96}
Cole S., Lacey C., 1996, \mnras,281,716

\bibitem[De Lucia \& Blaizot (2007)]{Del07}
De Lucia G., Blaizot J., 2007, \mnras, 375, 2

\bibitem[De Lucia \etal (2004)]{Del04}
De Lucia G., Kauffmann G., Springel V., White, S.~D.~M., 
Lanzoni B., Stoehr F., Tormen G., Yoshida N., 2004, \mnras, 348, 333

\bibitem[Fakhouri \& Ma (2008)]{FM08}
Fakhouri O., Ma C.~P., 2008, \mnras, 386, 577

\bibitem[Gao \etal (2004)]{Gao04a}
Gao L., White S.~D.~M., Jenkins A., Stoehr F., Springel V., 2004,
\mnras, 355, 819

\bibitem[Gao \etal (2005)]{Gao05}
Gao L., Springel V., White S.~D.~M., 2005, \mnras, 363, 66

\bibitem[Giocoli \etal (2008a)]{CG08a} 
Giocoli C., Torman G., van den Bosch F.~C., 2008a, \mnras, 386, 2135

\bibitem[Giocoli \etal (2008b)]{CG08b}
Giocoli C., Pieri L., Tormen G., \mnras, 2008b, 387, 689

\bibitem[Hopkins \etal (2006)]{Hop06}
Hopkins P.~F., Hernquist L., Cox T.~J., Di Matteo T., Robertson B., Springel V., 2006, \apjs, 163, 1

\bibitem[Kim \etal (2008)]{Kim08}
Kim J., Park C., Choi Y., 2008, \apj, 683, 123

\bibitem[Lacey \& Cole (1993)]{LC93}
Lacey C., Cole S., 1993, \mnras , 262, 627

\bibitem[Lemson \& Kauffmann (1999)]{Lem99}
Lemson G., Kauffmann G., 1999, \mnras, 302, 111

\bibitem[Li \etal (2007)]{Li07}
Li Y., Mo H.~J., van den Bosch F.~C., Lin W.~P., 2007, \mnras, 379, 689

\bibitem[Li \etal (2008)]{Li08}
Li Y., Mo H.~J., Gao L., 2008, \mnras, 389, 1419 

\bibitem[Lu \etal (2006)]{Lu06}
Lu Y., Mo H.~J., Katz N., Weinberg M.~D., 2006, \mnras, 368, 1931

\bibitem[Maller \etal (2006)]{Maller06}
Maller A.~H., Katz N., Keres D., Dave R., Weinberg D.~H., 2006, \apj, 647, 763

\bibitem[Mandelbaum \etal (2006)]{Mand06}
Mandelbaum R., Seljak U., Kauffmann G., Hirata C.~M., Brinkmann J., 2006, \mnras, 368, 715

\bibitem[McIntosh \etal (2008)]{Mc08}
McIntosh D.~H., Guo Y., Hertzberg J., Katz N., Mo H.~J., van den Bosch F.~C., Yang X., 
2008, \mnras, 388, 1537

\bibitem[Mo \& White (1996)]{Mo96}
Mo H.~J., White S.~D.~M., 1996, \mnras , 282, 347

\bibitem[Naab \& Burkert (2003)]{NB03}
Naab T., Burkert A., 2003, \apj, 597, 893

\bibitem[Navarro, Frenk \& White (1997)]{NFW}
Navarro J.~F., Frenk C.~S., White S.~D.~M., 1997, \apj, 490, 493

\bibitem[Sheth (2003)]{Sheth03}
Sheth R.~K., 2003, \mnras , 345, 1200

\bibitem[Sheth, Mo \& Torman (2001)]{SMT01}
Sheth R.~K., Mo H.~J., Tormen G., 2001, \mnras , 323, 1

\bibitem[Sheth \& Torman (1999)]{ST99}
Sheth R.~K., Tormen G., 1999, \mnras, 308, 119

\bibitem[Springel \etal (2005)]{Sp05}
Springel V. \etal, 2005, Nat., 435, 639

\bibitem[Springel \etal (2001)]{Sp01}
Springel V., White S.~D.~M., Tormen G., Kauffmann G., 2001, \mnras, 328, 726


\bibitem[Tinker \etal (2005)]{Tinker05}
Tinker J.~L., Weinberg D.~H., Zheng Z., Zehavi I., \apj, 2005, 631, 41

\bibitem[van den Bosch \etal (2005)]{vdb05}
van den Bosch F.~C., Tormen G., Ciocoli C., 2005, \mnras, 359, 1029

\bibitem[van den Bosch (2007)]{vdb07}
van den Bosch \etal, 2007, \mnras, 376, 841

\bibitem[Wang \etal (2006)]{Wang06}
Wang L., Li C., Kauffmann G., De Lucia G., 2006, \mnras, 371, 537

\bibitem[Wechsler \etal (2002)]{Wec02}
Wechsler R.~H., Bullock J.~S., Primack J.~R., Kravtsov A.~V., Dekel A.,
2002, \apj, 568, 52

\bibitem[Wetzel \etal (2008)]{Wet08}
Wetzel A.~.R., Cohn J.~D., White M., 2008, preprint (astro-ph/0810.2537) 

\bibitem[Yang \etal (2003)]{Yang03}
Yang X.~H., Mo H.~J., van den Bosch F.~C., 2003, \mnras, 339, 1057

\bibitem[Yang \etal (2009)]{Yang09}
Yang X.~H., Mo H.~J., van den Bosch F.~C., 2009, \apj, 693, 830

\bibitem[Zhao \etal (2003)]{Zhao03a}
Zhao D.~H., Mo H.~J., Jing Y.~P., B\"orner G., 2003, \mnras , 339, 12

\bibitem[Zheng \etal (2005)]{Zheng05}
Zheng Z. \etal, 2005, \apj, 633, 791

\end{thebibliography}
\end{document}